\documentclass[sort&compress]{elsarticle}
\usepackage{lineno}
\usepackage{amssymb}
\usepackage{amsmath,epsfig}
\usepackage[latin1]{inputenc}
\usepackage{graphicx}
\usepackage[english]{babel}
\usepackage{xspace}
\usepackage{subfigure}
\usepackage{float}
\usepackage{multirow}
\usepackage{hyperref}
\usepackage{placeins}
\usepackage[percent]{overpic}
\usepackage{eurosym}
\usepackage{tablefootnote}
\usepackage{hyperref}
\usepackage{array}
\newcolumntype{L}[1]{>{\raggedright\let\newline\\\arraybackslash\hspace{0pt}}m{#1}}
\newcolumntype{C}[1]{>{\centering\let\newline\\\arraybackslash\hspace{0pt}}m{#1}}
\newcolumntype{R}[1]{>{\raggedleft\let\newline\\\arraybackslash\hspace{0pt}}m{#1}}

\usepackage{color}

\makeatletter
\def\ps@pprintTitle{%
 \let\@oddhead\@empty
 \let\@evenhead\@empty
 \def\@oddfoot{}%
 \let\@evenfoot\@oddfoot}
\makeatother
\journal{Nuclear Instruments and Methods in Physics Research }

\begin{document}

\begin{frontmatter}

\title{Fast timing detectors with applications in cosmic ray physics and medical science}
\author[add1]{C.~Royon}
\ead{christophe.royon@cern.ch}
\author[add2]{F.~Gautier}
\ead{florian.gautier.borrallo@gmail.com}

\address[add1]{University of Kansas, Lawrence, USA.}

\begin{abstract}
We  discuss the use of Low Gain Avalanche (LGAD) silicon detectors for two specific applications, namely  measuring cosmic rays in space in collaboration with NASA] and beam properties and received doses for patients undergoing cancer treatment in flash beam therapy. 
\end{abstract}


\end{frontmatter}


\section{Introduction: LGAD silicon detectors}

In this report, we will discuss the use of Low Gain Avalanche (LGAD) silicon detectors. These detectors were initially developed
for particle timing measurements at the Large Hadron Collider at CERN. We will focus on two specific applications for LGAD detectors. The first application involves measuring cosmic rays in space, and this research is conducted in collaboration with NASA. The second application pertains to the measurement of beam properties and received doses for patients undergoing cancer treatment. This is especially relevant in the context of flash beam therapy.

In order to obtain the largest possible signal from a silicon detector, we need large velocity $v$ (which means a fast detector), large fields $E_W$  and large pads to have a uniform field, and lots of charge $q$ following the Ramo's theorem $I \sim q v E_w$. If we want to measure many particles entering in our detector as a function of time (for instance cosmic rays in space and electron/protons crossing our detector during flash beam cancer therapy), it is clear that we need a short signal in order to avoid pile up between different particles. For this sake, we chose ultra fast silicon detectors (UFSDs) that seemed more adapted to our problems that lead to typical signal of a few nanoseconds. These detectors are already used at the LHC for instance in the TOTEM experiment in order to measure the proton time-of-flight with precisions close to 20 ps per layer. They will also be used to measure time-of-flights of all particles (in the central and/or the forward parts of the detectors) in the ATLAS and CMS collaborations for the high luminosity LHC~\cite{nicola,delagnes}.

In order to measure the signal, the first step is usually to amplify the signal using an amplifier designed at the University of Kansas using standard components. 
The second step is to perform a fast sampling of the signal and measure for instance 256 points on the rising part of the signal or on the full signal. It allows to measure simultaneously the time-of-flight of the particle with a precision of 10 to 15 picoseconds (by measuring the starting point of the signal in the LGAD), the pulse amplitude and shape. This will be the basics information used for all applications.

\begin{figure}[!h]
\centerline{%
\includegraphics[width=0.7\textwidth]{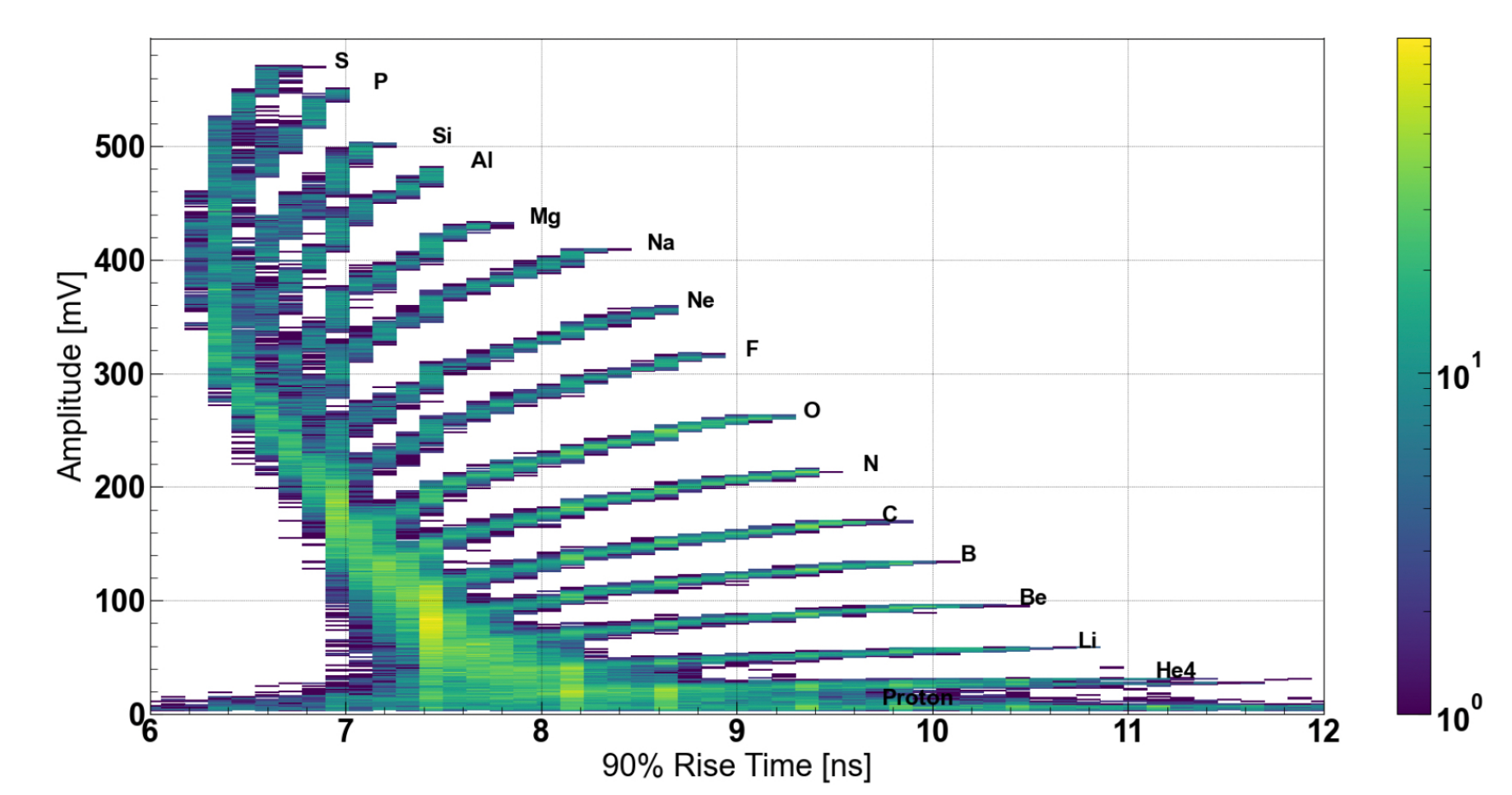}}
\caption{Signal amplitude versus 90\% of rise time using ultra-fast silicon detectors. This shows the possibility to identify the type of particles in cosmic rays in space since the curves do not overlap for most of the phase space. }
\label{fig1}
\end{figure}

\begin{figure}[!h]
\centerline{%
\includegraphics[width=0.7\textwidth]{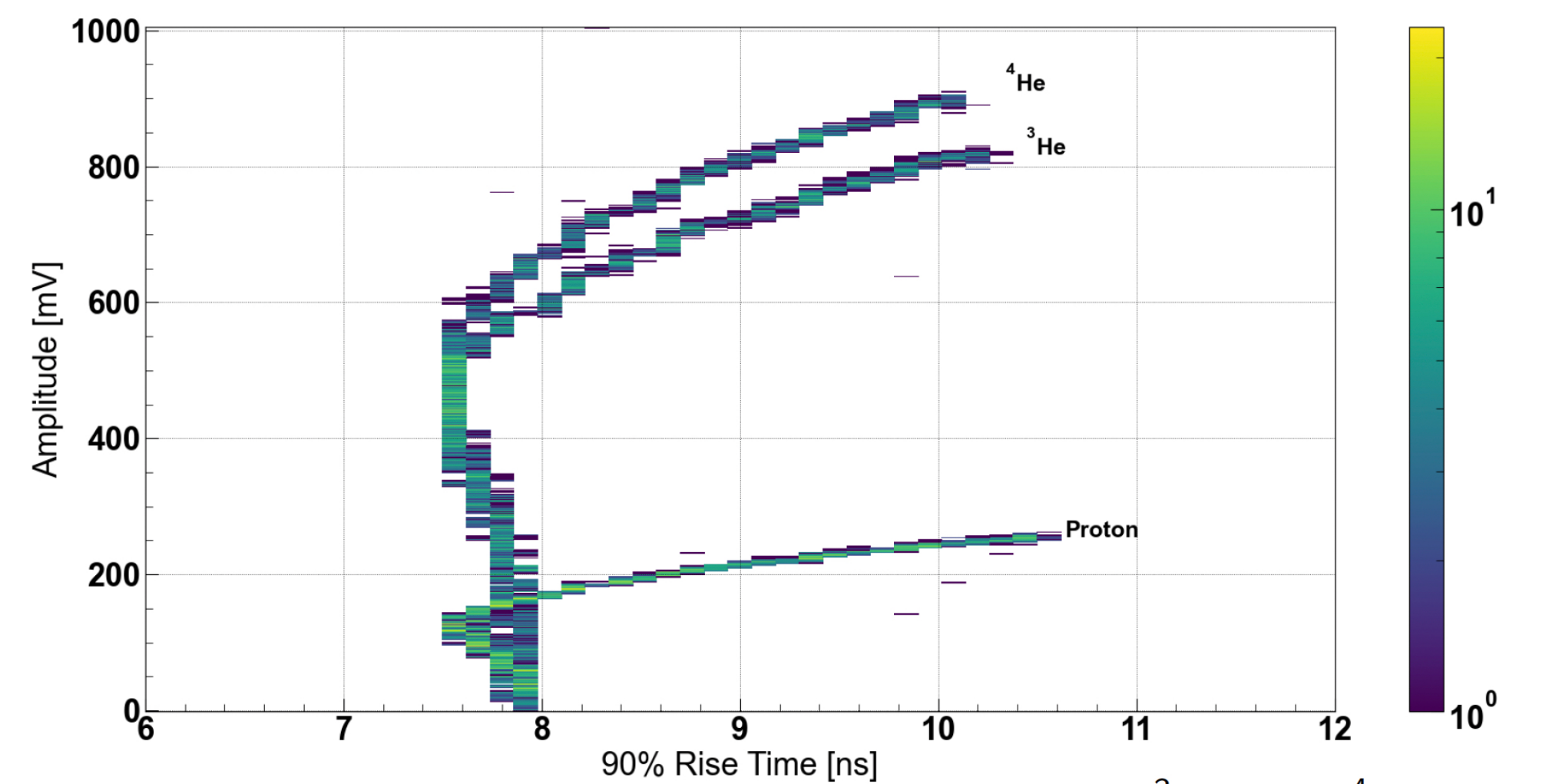}}
\caption{Signal amplitude versus 90\% of rise time  using ultra-fast silicon detectors showing the possibility to identify the different isotopes for He. This is crucial to distinguish between different models of particle emissions by the sun.}
\label{fig2}
\end{figure}

\section{Measuring cosmic rays in space: the AGILE project}

The goals of AGILE (Advanced Energetic Ion Electron Telescope) is to build a compact, low power and low cost instrument for characterization of solar energetic and anomalous cosmic ray particles originating from the sun and outer space in collaboration with NASA~\cite{agile}. The measurement should be performed in space.  We focus on Ions (H-Fe), with energies in the range 1-100 MeV/nucleon and also on electrons with energies between 1 and 10 MeV. AGILE will thus perform robust real-time particle identification and
energy measurement in space. The solution is to use multiple layers of LGAD detectors (with or without absorbers) and 
to measure the signal in the stopping layer. AGILE could be 
upgraded to higher energy ranges in the future by using more layers of detectors or absorbers.

As a prototype to be sent into space, we chose to build a simplified version of AGILE with three layers of UFSD. The measurement of the signal is performed in each layer using the fast sampling technique. This is the first time that fast sampling will be performed in space. The idea is then to extract the ion identification ($p$, $He$, $Au$, $Pb$, etc) and their energies by measuring the amplitude and duration of the signal. In order to determine the accuracy of such measurement, we performed a full simulation starting from the particle simulation using GEANT4~\cite{geant4} to the interaction with the detector using Weightfield~\cite{weightfield} and the extraction of the signal using LTspice~\cite{ltspice}.

\begin{figure}[!h]
\centerline{%
\includegraphics[width=0.7\textwidth]{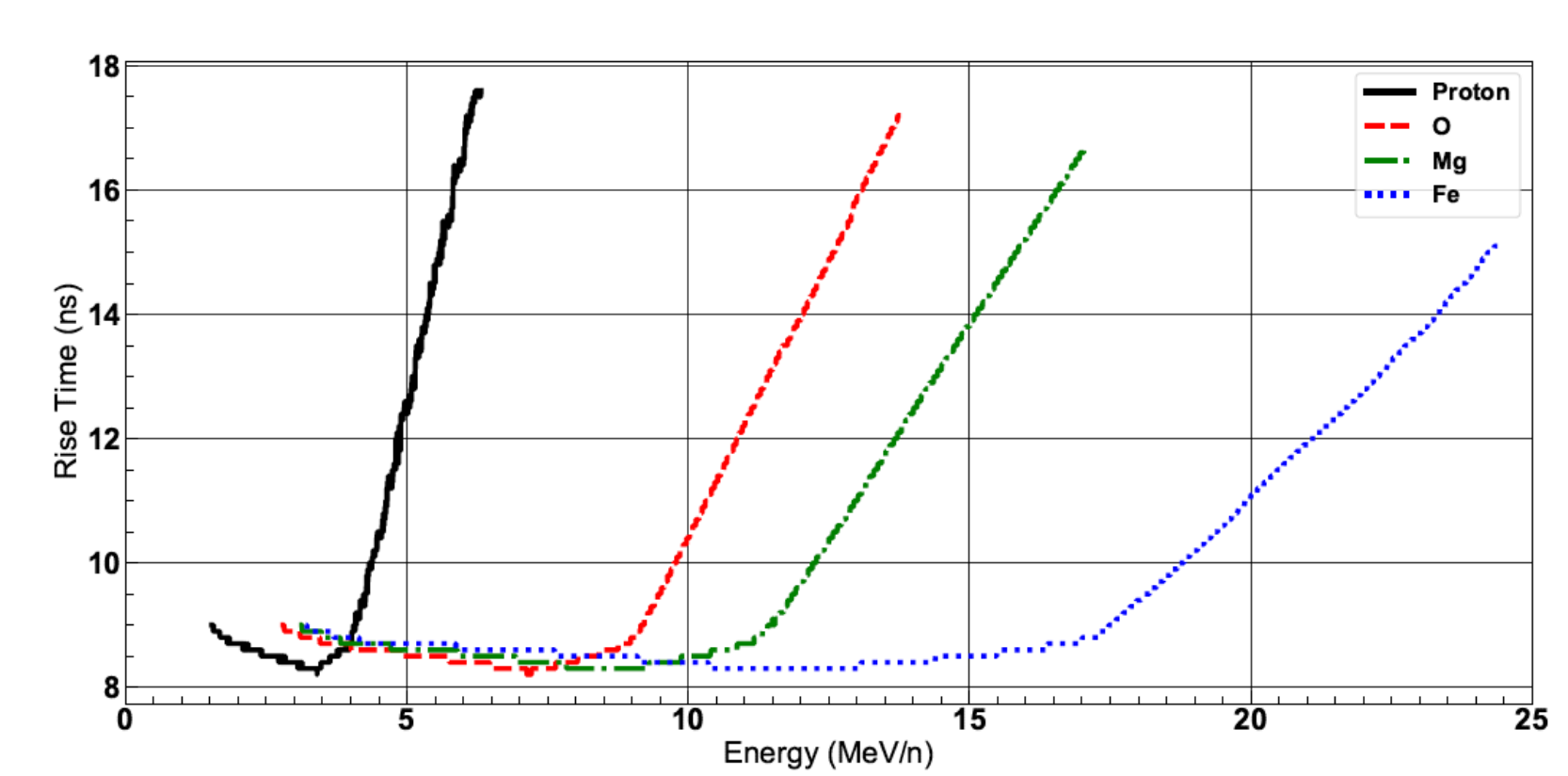}}
\caption{Rise time vs energy  using ultra-fast silicon detectors. Once the type of particle is identified (see Fig.~\ref{fig1}) this shows the possibility of measuring the particle energy since the different curves do not overlap.}
\label{fig3}
\end{figure}

From the simulation, we identified the maximum amplitude of the signal in the stopping layer and the time to reach 90\% of the maximum of the signal (rise time) as key elements to perform particle identification and energy measurements. The maximum amplitude vs the rise time for p-Fe ions stopping in the
detector is shown in Fig~\ref{fig1}. This allows to obtain particle identification (the $Z$ value)  since the different curves do not overlap for most values of rise time. There is still some inefficiencies for small rise time values where curves overlap (when the signals are too short).  It is also worth noticing that we have large differences between a typical LGAD signal originating from a minimum ionizing electron, 100 MeV proton or a Pb ion stopping in the detector, as an example. It  leads to signals of about 3. 10$^{7}$, 2. 10$^{-6}$ and 5. 10$^{-3}$ A, respectively. These large ranges of signal amplitudes  were the reason why we developed two different amplification chains on our readout card allowing to measure lower signal with better accuracy. In Fig.~\ref{fig2}, we show the maximum amplitude vs rise time for protons, $^3$He, and $^4$He ions. We see that we can distinguish between the different He isotopes which is quite fundamental for cosmic ray models. In Fig.~\ref{fig3}, we display the rise time vs energy showing the possibility of measuring the particle energy once the type of the particle is identified (using Fig.~\ref{fig1}. 

\begin{figure*}
\centering
\includegraphics[width=0.45\textwidth]{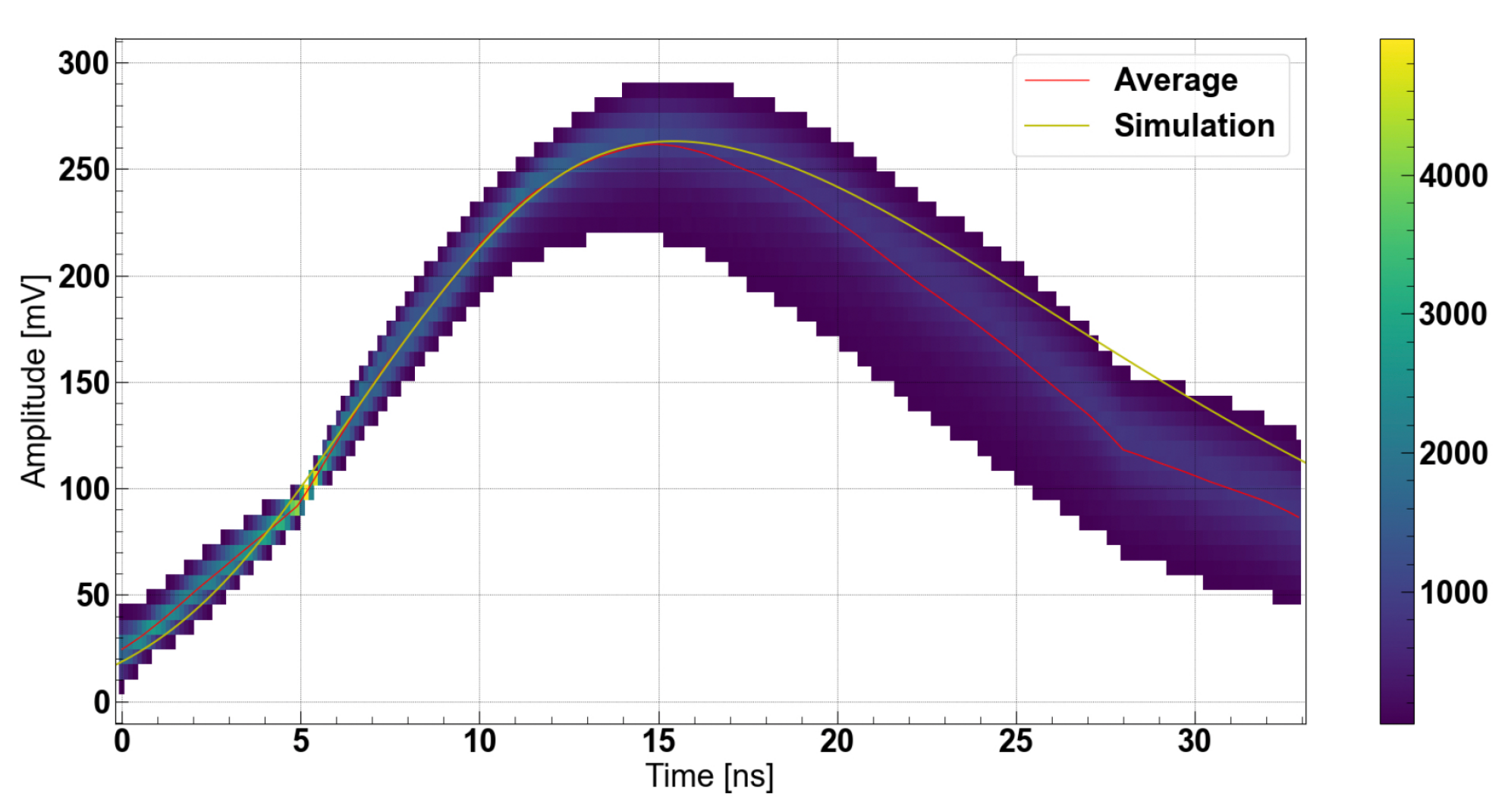}
\includegraphics[width=0.45\textwidth]{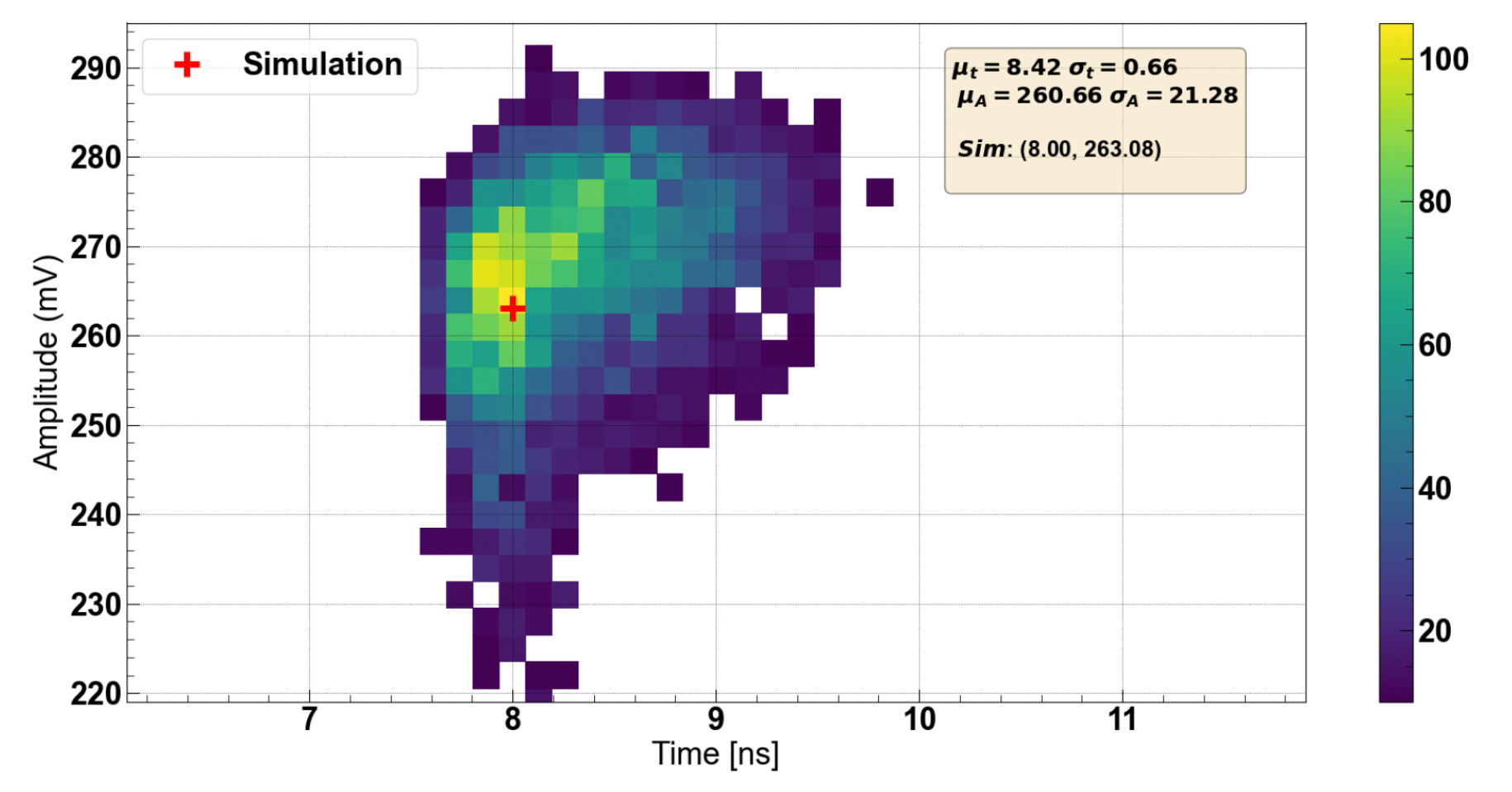}
\caption{Left: amplitude vs time distribution as measured by AGILE for alpha particles of about 5.5 MeV  (Americium-241 radioactive source) compared to simulation. Right: maximum amplitude vs 90\% Rise time for alpha particles measured by
AGILE in the laboratory at KU compared to simulated value. We notice the good agreement between simulation and data.}
\label{fig4}
\end{figure*}

The launch of AGILE by NASA is foreseen by the end of 2023. Beam tests to check in depth the results of the simulation with respect to data occurred at Brookhaven National Laboratory at the end of May. Some tests were already performed at the University of Kansas and at NASA using a radioactive source (Americium-241). In Fig.~\ref{fig4}, we show on the left the amplitude vs time distributions as measured by AGILE for alpha particles originating from the Americium-241 radioactive source compared to simulation On the right, we display the maximum amplitude vs the 90\% Rise time.  A good agreement is found between measurement and simulation, that will be further tested using a full beam test using different kinds of particles (p, H to Fe).

The energy resolution due to statistical fluctuations of the energy
deposition, the number of charge carriers and the electronics noise were estimated to be between 1 and 3\% using the simulation for oxygen as an example. This is well below the requirements for AGILE of 
$\Delta E/E <$ 30\%. Lighter heavy ions lead to similar performances while heavier heavy ions such as Au or Pb show performances better than 1\%. 

The AGILE acceptance is shown in Fig.~\ref{fig5}, left, where we display the energy vs ion charge (Z) possible measurements from the simulation. The blue regions correspond to the domains where AGILE
can identify the particles and measure their energy
while the hatched orange regions correspond to the region
where discrimination is not possible.  This means an inefficiency of our method  (because there is not enough separation in the maximum amplitude and/or rise time between the different ions).  This inefficiency translates into regions of energies where measurements are not possible.
This is fine for the physics measurements that we would like to perform and it can be further improved by using for instance different widths of LGADs. 

We also studied the dependence of our results on the angular acceptance of AGILE. A field of view of 40 degrees (20 degrees of half angle) is fine for all ion determinations while getting a slightly wider spread of the curves maximum amplitude vs rise time. A wider angle makes discrimination more challenging. This is why we limited the angle aperture of AGILE by about 40 degrees. We also studied the variation of the signal characteristics (maximum amplitude and rise time) as a function of  temperature between (-40) and (+40) degrees. The typical variations are less than 20\% for the amplitude and 7\% for the rise time.  Given the fact that the dependence as a function of temperature  is  linear, this
can be easily corrected on-board since temperature will be measured with an accuracy of about 0.1 degree. 

A view of the AGILE mockup with three layers of LGADs as it will be sent into space is shown in Fig.~\ref{fig5}, right. Future improvements of AGILE will involve more layers of LGADs and also more advanced techniques (such as machine learning using the 256 points of measurement using the fast sampling method) in order to obtain even better efficiencies on ion type determination and energy measurements. The idea of these further developments would be to have a network
of cube satellites that can measure cosmic ray events simultaneously and also to perform a detailed measurement of radiation between the Earth and Mars before sending astronauts to Mars.

\begin{figure*}
\centering
\includegraphics[width=0.45\textwidth]{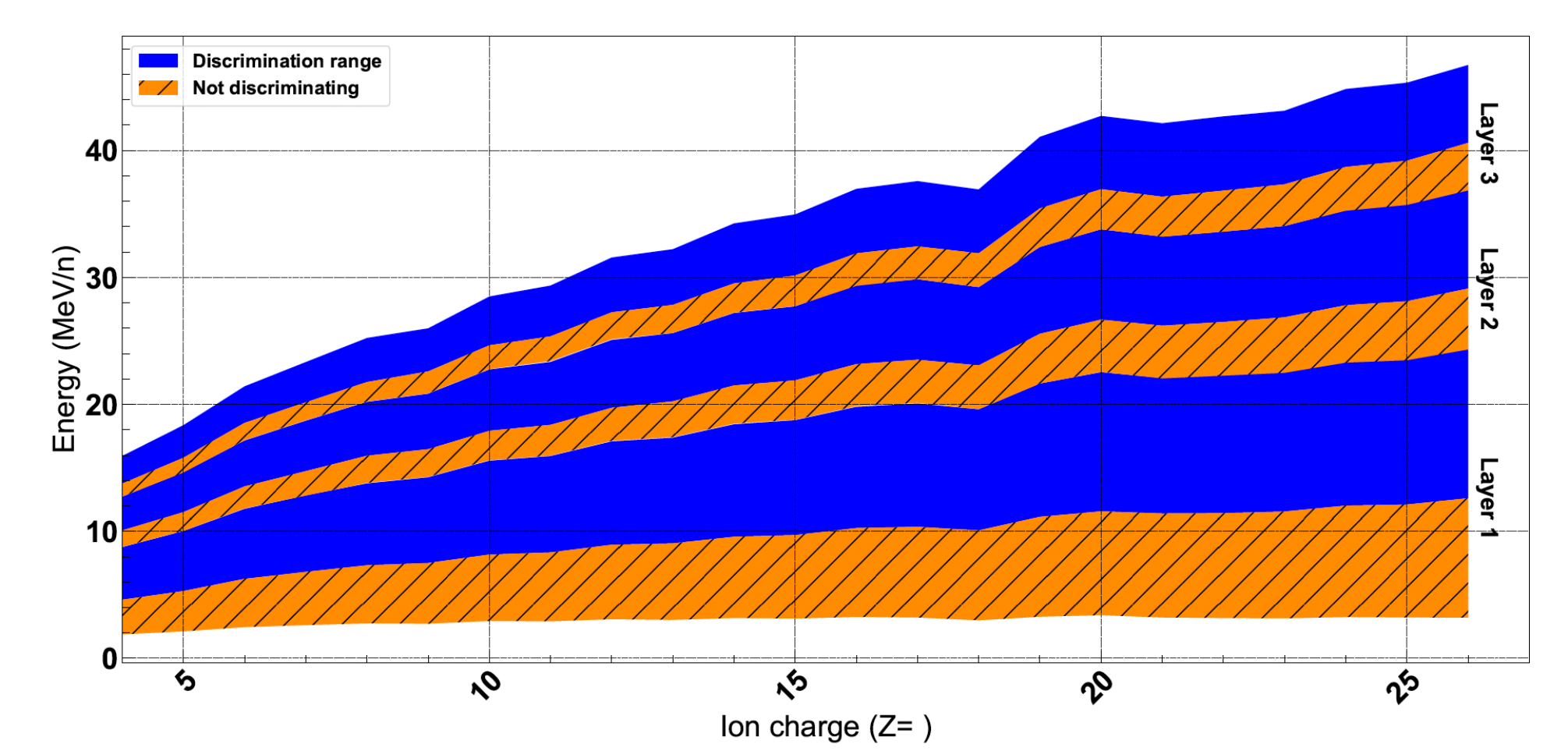}
\includegraphics[width=0.45\textwidth]{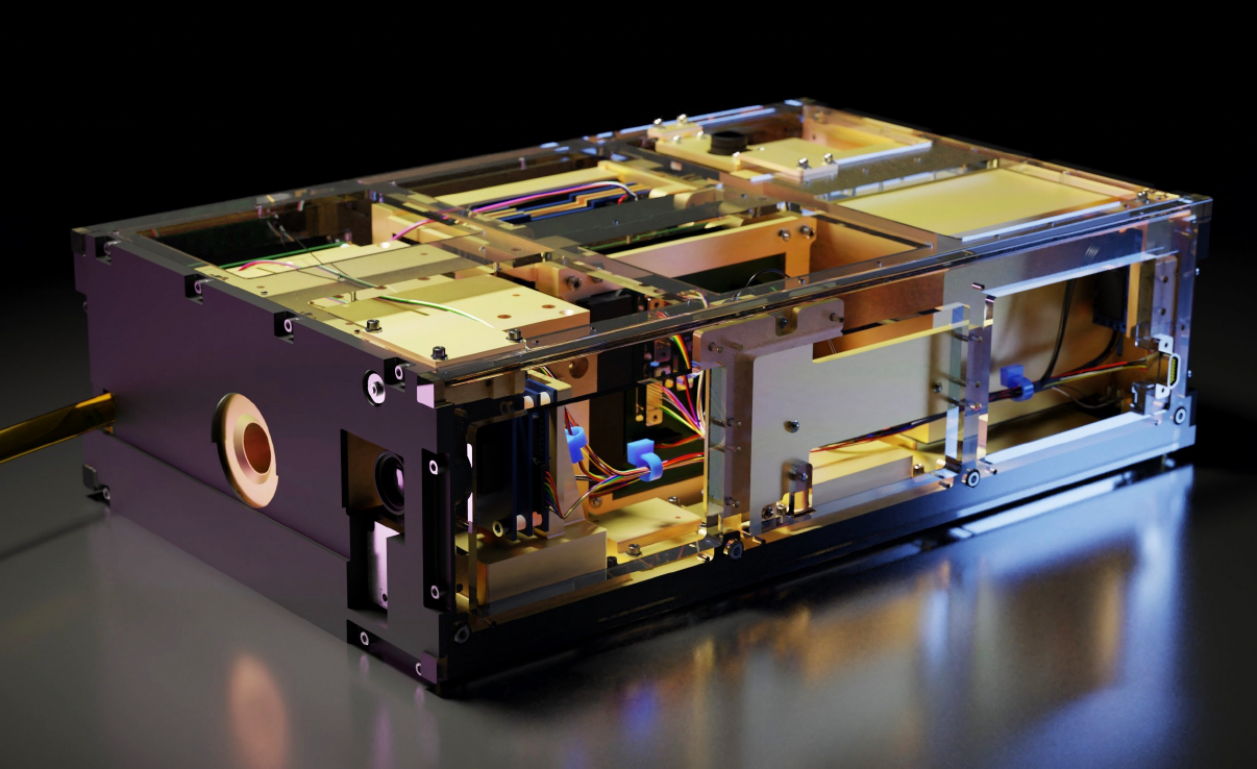}
\caption{Right: Performance of AGILE in the Energy vs ion charge plane. We note that AGILE can perform particle identification and energy measurement for a large domain of the phase space. This can be further improved using advanced analysis methods such as machine learning. Left: Mockup of AGILE as it will be sent into space.}
\label{fig4b}
\end{figure*}

\section{Medical application: measuring beam quality and doses in flash beam therapy}

The idea is to perform precise and instantaneous measurements of doses during cancer treatment (especially for flash proton beam therapy), and also to check online the quality of the machine. It needs indeed delivering the beam intensity and energy that it is supposed to deliver. The goal is thus to develop a fast and efficient detector to count the particles up to a high rate, allowing a very precise instantaneous dose measurement without the  need of calibration and with a high granularity ($mm^2$). Many different methods are used traditionally for dose measurements in cancer treatment, starting from ion chambers used in hospitals. Many new methods exist that can measure the radiation dose information in vivo on patients with Cerenkov or radiation induced acoustic imaging. We choose to concentrate on UFSDs since we want to use these detectors for flash beam therapy where the beam intensity is large. We thus need a fast detector that allows to count the number of particles (protons or electrons) that go across and that is also radiation hard.

\begin{figure*}
\centering
\includegraphics[width=0.7\textwidth]{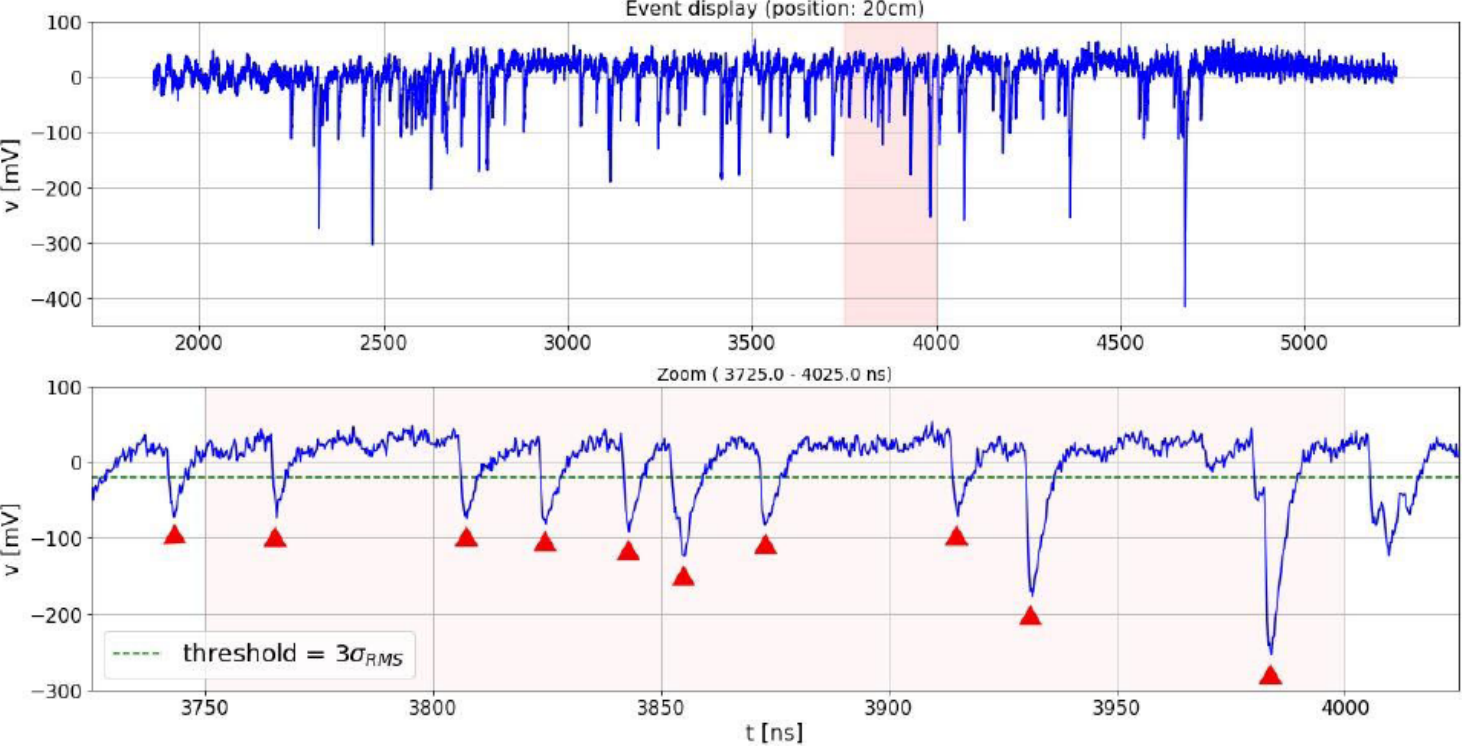}
\caption{Signal from the UFSD allowing to measure doses almost instantaneously by counting the spikes in signal due to the particles crossing the detector. The bottom plot shows the zoomed region in red from the top plot. The red triangles show the particles that lead to a trigger of our detector.}
\label{fig5}
\end{figure*}

We put our LGAD silicon detectors and readout system in an ELEKTATM Precise  Linac electron beam, with a pulse length  of about 3.2 $\mu$s long. This machine was used in the past for photon therapy at St Luke Hospital, Dublin, Ireland~\cite{medical}. Each pulse sequence contains thousands of 30 ps sub-pulses separated by
350 ps (with a frequency of 2.858 GHz). The electron beam has an energy between 4 and 18 MeV,
with dose rates up to 600MU/min, and pulse repetition frequency of 200 Hz. Our LGAD detector was mounted on a moving support to provide the monitoring of the beam as a function of
its location. A Neodymium N40 permanent magnet was located 12 cm
below the collimator to separate the charged and neutral particles from the beam.  
Using the good timing resolution of LGADs and the short duration of the signal, we were able to trigger and count on the particles entering our detectors as shown in Fig.~\ref{fig5}. We can see that we can count with high precision the number of particles that are identified as spikes in the LGAD. 

\begin{figure*}
\centering
\includegraphics[width=0.45\textwidth]{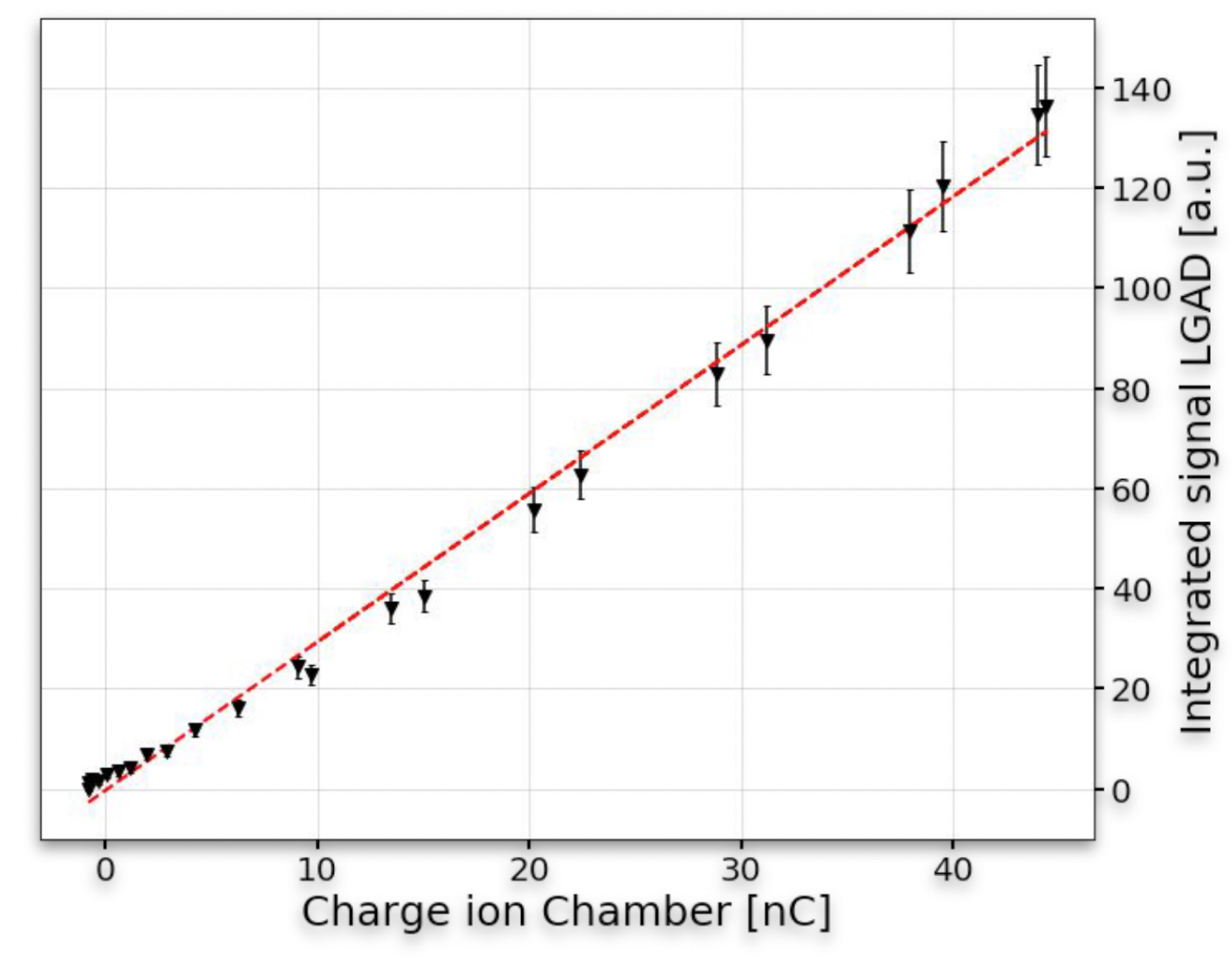}
\includegraphics[width=0.45\textwidth]{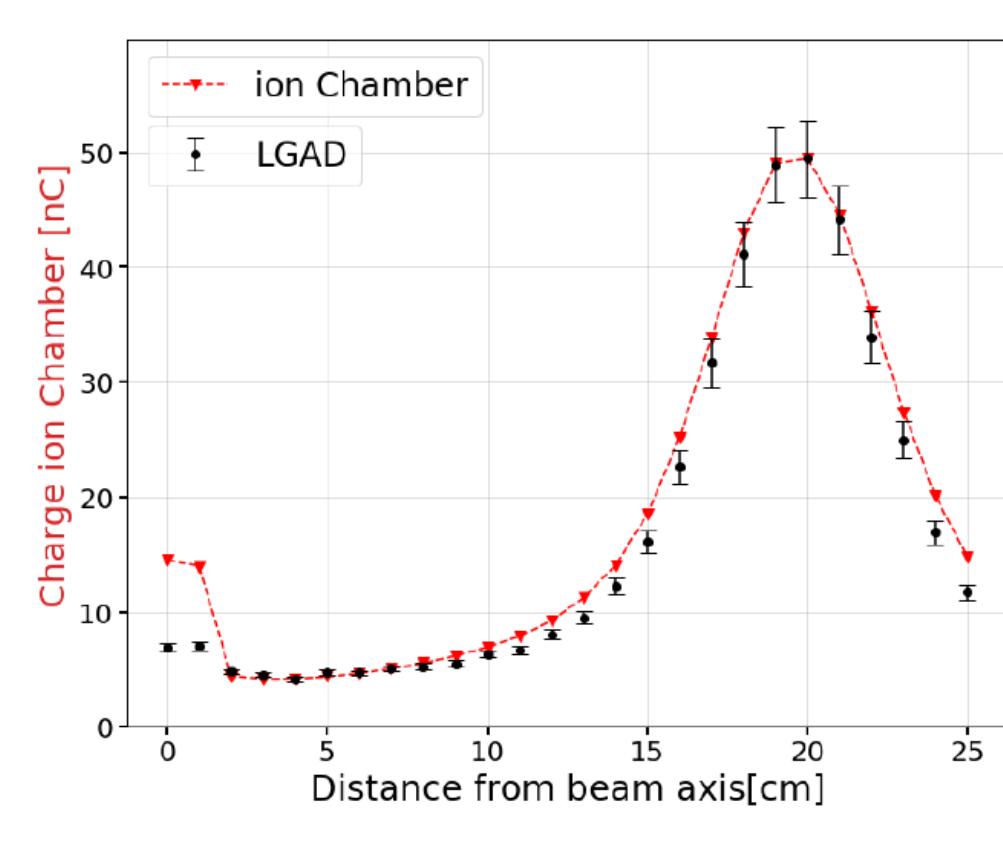}
\caption{Left: Correlation between the LGAD signal and the ion chamber results that show a good agreement between bothe measurements. The ion chamber can only perform measurements every few seconds while LGAD has a time resolution of about 20 to 25 ps. Right: Comparison between the measurement of LGAD and ion chamber as a function of distance from the beam and we not again a good agreement between both measurements.}
\label{fig6}
\end{figure*}

\begin{figure*}
\centering
\includegraphics[width=0.7\textwidth]{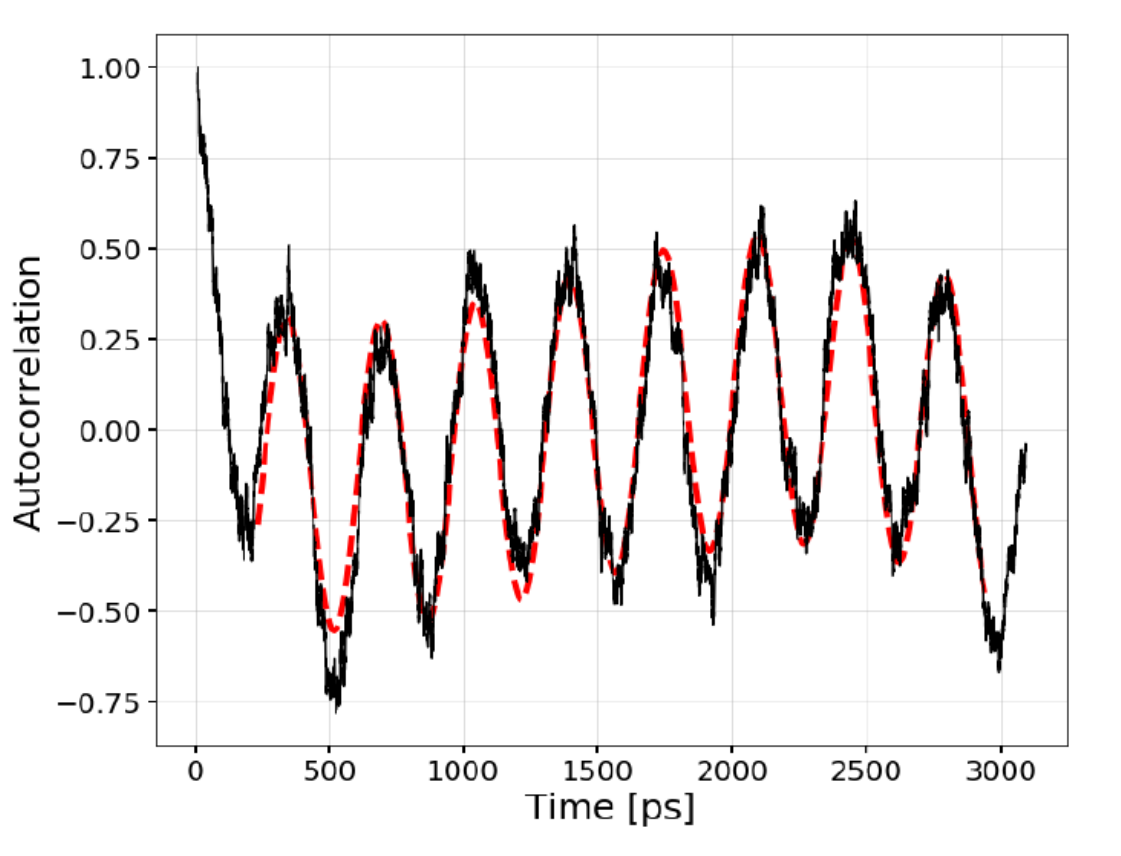}
\caption{Beam structure analysis in Dublin hospital using the UFSD detector. It was possible to see the beam structure for the first time benefitting from the short signals of UFSDs.}
\label{fig7}
\end{figure*}

The measurement of the charge deposited in the LGAD detector compared to standard measurement using an ion chamber is shown in Fig.~\ref{fig6}, left. A very  good correlation is found between both measurements. This shows that we are able to reproduce the dose measurement from the ion chamber. In Fig.~\ref{fig6}, right, we also display the comparison between the LGAD and ion chamber results as a function of distance from the center of the beam. We see a very good agreement except at low distances. This is due to the lower efficiencies of LGADs to photons that represent the majority of particles going straight, whereas the electrons are bent by the magnet and thus appear at larger distances.
In addition, the LGAD detectors allow to measure the beam structure as shown in Fig.~\ref{fig7} for the first time at the Dublin hospital. The good timing resolution of LGADs of 15 to 20 picoseconds allow to see the periodicity of the beam of about $\sim$330 ps, which is not possible using an ion chamber with an integration time of a few seconds. This method allows to  measure single particles from the beam, which is fundamental to measure instantaneous doses for high intensity proton/electron therapy for instance and to check online the quality of the machine.

We recently got a project approved by NIH with the goals of designing and building a UFSD dosimetry prototype
using the fast sampling method, and of testing 
this prototype in a proton/electron flash beam facility at the University of Kansas and the University of Washington in St Louis. The longer term goals will be to move from  a 8/16 channel readout board to a 100 or 1000 channel board that could have direct commercial applications to measure the machine quality and doses received by patients in flash beam therapy treatments.

\section{Conclusion}
Fast timing and LGAD detectors were originally developed for high energy physics at CERN. We developed at KU a readout electronics card using standard components for different applications, the idea being to
reconstruct the full spectrum of a given signal using the fast sampling technique.
The first application that we discussed is to measure cosmic ray particles (both identify the type of particles and measure their energies) in a cube satellite in collaboration with NASA using the Bragg peak technique. Results are already very promising using an Americium-241 source.
The second application deals with the instantaneous measurement of beam quality and  doses in flash beam therapy  during cancer treatment with high accuracy by counting the number of particles. 
The first results obtained at the hospital in Dublin show  that we indeed see the beam structure benefitting from the fast signal properties of LGADs and that we are able to measure the doses with high precision. The next step will be to put our detector in a flash beam available at the hospitals at the University of Kansas and the University of St Louis. Further developments will include new electronics, detector and software optimizations in order to reconstruct the signal when a few protons and electrons signals overlap in our detector.

\section*{Acknowledments}
The authors thank A. Greeley, T. Isidori, S. G. Kanekal, G. Legras, P. McCavana, B. McClean, R. McNulty, N. Minafra, A. Novikov, N. Raab, L. Rock, Q. Schiller for useful collaboration.
This work is supported by Heliophysics Technology and Instrument Development for Science (HTIDS) ITD and LNAPP program part of NASA Research Announcement (NRA) NNH18ZDA001N-HTIDS for Research Opportunities in Space and Earth Science 2018 (ROSES-2018).

\end{document}